# On the relationship between interdisciplinarity and impact: different modalities of interdisciplinarity lead to different types of impact

Jordi Molas-Gallart[1,*], Ismael Rafols[1,2] and Puay Tang[2]

[1] *Ingenio* (CSIC-UPV), Universitat Politècnica de València, València
[2] SPRU – Science and Technology Policy Research, University of Sussex, Brighton
* Corresponding author: jormoga@ingenio.upv.es

*Version of 28th July 2014*

**Abstract**
There is increasing interest among funding agencies to understand how they can best contribute to enhancing the socio-economic impact of research. Interdisciplinarity is often presented as a research mode that can facilitate impact but there exist a limited number of analytical studies that have attempted to examine whether or how interdisciplinarity can affect the societal relevance of research. We investigate fifteen Social Sciences research investments in the UK to examine how they have achieved impact. We analyse research drivers, cognitive distances, degree of integration, collaborative practices, stakeholder engagement and the type of impact generated. The analysis suggests that interdisciplinarity cannot be associated with a single type of impact mechanism. Also, interdisciplinarity is neither a sufficient nor a necessary condition for achieving societal relevance and impact. However, we identify a specific modality -- "long-range" interdisciplinarity, which appears more likely to be associated with societal impact because of its focused problem-orientation and its strong interaction with stakeholders.

## 1. Introduction

Interdisciplinary Research (IDR) is often described in science policy discourse as necessary so that science has a socio-economic impact. Indeed, it is generally claimed that interdisciplinary research is particularly relevant for addressing social problems and for fostering innovation (Chavarro, Tang and Rafols, 2014). This articles aims to investigate whether and how interdisciplinarity has an influence on impact by examining a number of projects funded by the UK's Economic and Social Research Council (ESRC).

IDR has become a broad brush to refer to a wide variety of research strategies and practices. Science policy and research support programmes are increasingly paying attention to IDR and with this increased attention more research initiatives are claiming to be interdisciplinary (Braun and Schubert, 2003). As it is often the case, conceptual imprecision and ambiguity follow when a concept becomes fashionable. To understand how IDR is linked with impact and to help the development of Council strategies to deal with it, we first need to understand what IDR is in practice.

There is abundant literature addressing the nature of IDR and associated concepts (multidisciplinarity, transdisciplinarity,…) (Wagner et al., 2011). Most of this interest is associated with views that stress the importance of new forms of research bridging the institutional boundaries that have overtime set scientific research within differentiated areas (disciplines) with their own epistemologies, validity criteria and associated institution-building tools (journals, conferences, university departments, degrees, professional associations). Bridging, combining or transcending these disciplines boundaries is seen as a means of sparking creativity in science, supporting innovation and addressing pressing social needs (Zierhofer and Burger, 2007; Jacobs and Frickel, 2009).

Yet, although there is much anecdotal evidence supporting the claims that IDR often addresses social and economic problems and helps in providing solutions, findings on the relationship between IDR and impact are not systematic and have been questioned (Jacobs and Frickel, 2009). For example, some studies do not find that interdisciplinarity influences the success of firms founded by academic teams (Muller, 2009), while other studies correlate interdisciplinary practices with the intensity of university-industry interactions (Carayol and Thi, 2005). This variety of results suggests that the different types of IDR and the various processes through which research is defined, funded, conducted and applied, have an effect on the extent and type of social impact. Although IDR is often associated with problem-orientation (because societal problems seldom conform to disciplinary boundaries) and with research that entails interactions beyond academia, one can also think of IDR that does not address societal issues and does not interact with their potential non-academic beneficiaries, for example in areas such as biophysics. Further, as we will see below, the impact processes through which IDR can yield socio-economic benefits need not be unique to IDR. The relationship between IDR and impact is therefore complex and we should not assume it can be described through simple models of general applicability that simply differentiate IDR from disciplinary research.

Therefore, our study needs to address both the way in which IDR is interpreted and practised by project researchers and stakeholders, and the processes and mechanisms through which IDR in these projects leads to socio-economic impacts. Following our previous studies on research impact, we will focus on processes; we understand that a good way to trace the links between a social or economic impact and the research that has contributed to such impact, is through an analysis of the interactions between researchers and non-academic stakeholders. When a direct contact leads to an effort by a stakeholder to engage with the research or its results, we say that a "productive interaction" has taken place. When the productive interaction leads stakeholders to do things differently or to do new things, we say that the research has had an impact (Molas-Gallart and Tang 2011; Spaapen and van Drooge 2011).

The article is organised as follows. First we describe the methodological approach and data used. Second we look at the rationale and objectives of the programmes and projects used as case studies. Third, we describe the type of IDR research that was conducted, the forms of stakeholder engagement and their associated impacts. Fourth, we report the perceptions of the participants on the nature and effects of IDR. The last two sections discuss the findings and the policy implications.

## 2. Methodological approach

This exploratory study, commissioned by the ESRC, is mainly based on existing reports on a selection of ESRC investments and a limited number of interviews with stakeholders and researchers in two of the selected projects. We studied ESRC-funded programmes, centres and projects (to which we refer to as ESRC investments) implemented over the past 10 years. The investments were selected because they explicitly included interdisciplinary elements in their research design, although these were interpreted differently across investments. An anonymised description of the investments selected can be found in the Annex to this report and include three Research Programmes, a Research Centre, and eleven Research Projects.

- **Research Programmes** involve several projects under the umbrella of a common research problem; they take place over a period of time that commonly stretches to five or ten years and usually involve more than ten projects. When taking the Programme as a unit of analysis we were mainly concerned with the activities that the Programme, under the guidance of its Director, attempts to carry out; for instance, the Programme may seek projects following interdisciplinary strategies such as inter-organizational collaborations or training activities, or may explicitly support IDR using tools like training or dissemination activities designed for this purpose.
- **Research Centres** are large investments (typically around £5 million) with a duration similar to Research Programmes, but with the research activities carried out directly by the groups receiving the funding. They are therefore more centralised than Programmes, with a common schedule of activities and an amount of resources that allows for the medium-term engagement of stakeholder communities. Centre Directors will typically develop a dissemination and engagement activity, and the Centre will have a pre-defined set of research objectives and associated research methods.
- **Research Projects** are smaller in size, but there is still substantial variability in their size and scope. The Projects reviewed for this analysis had budgets that varied from little over £100,000 to well over £1,000,000. Projects are typically research activities with a specific research objective and method; but among the projects we reviewed we found also investments that had as its main objective the generation of research networks around a specific research topic. The ways in which a small empirical research project and a networking activity can pursue interdisciplinary research objectives are likely to be different.

The investments are referred to throughout the report by their acronyms or abbreviations as listed in the Annex.

We used two data sources. The core of the analysis is based on previous reports held by the ESRC: End of Award, Impact reports, Rapporteurs' comments, and Impact Studies carried out by external experts. These provided information on the activities conducted, their rationale, and the results obtained. They were used to assess how the projects refer to IDR, and the forms and practices of interdisciplinary research they implemented. This study quotes extensively from these reports.

These written reports, however, do not present information about the ways in which non-academic stakeholders view IDR and on how the research results have been used and applied once the project itself has concluded. The study was therefore complemented with a set of interviews with researchers and stakeholders from projects P_Sleep and P_Water (see annex for brief description of project).

**3. IDR and Impact-Performing IDR**

IDR can mean different things to different people. The concept encompasses many diverse ways of combining different bodies of knowledge. We distinguish four main aspects in which important differences emerge: (a) the primary objectives driving the research; (b) the distance between the bodies of knowledge that IDR bridges (cognitive distance); (c) the extent of integration among these bodies of knowledge that IDR brings about; and (d) the practices by which researchers conduct IDR.

### 3.1 Research drivers

All the investments studied for this report have an applied side to their design or purpose. In most of the investments we reviewed, specific societal problems were driving the research. Yet, the way the problems were laid out and related to the research objectives varied substantially across projects. We differentiate three main different ways in which "problem-orientation" appears in the projects analysed.

1. **Projects addressing a specific and narrowly bounded problem.** For instance, P_Sleep aimed to develop solutions to deal with the problem of lack of sleep among older people. P_Water investigated the "generative role" of knowledge controversies to develop the societal capacity to deal more effectively with the uncertainty of flooding. These projects sought to generate solutions for these problems, the solution of which would typically include an engineering or a software component. In this context IDR appears as the direct result of the need to combine different forms of knowledge to generate specific solutions.
2. **Initiatives addressing a problematic area (a "challenge"), which can generate a variety of specific problems related to the challenge.** When dealing with a broader set of policy issues, the *research aims to be policy-relevant, but not necessarily solution-oriented*. For instance, it is possible that a main concern of the project can be theoretical or methodological but with the explicit intention of making the resulting body of theory better suited to tackle practical problems. For instance, P_Health was set up as a research network to combine two different bodies of theory (the study of the social determinants of ill-health and mortality with the "Capabilities Approach"), with the explicit aim of obtaining "prescriptive power" by "laying the conceptual groundwork for building a theory of health causation and distribution that is applicable across human societies". P_Nature was set up as a collaborative network of researchers and public sector/policy representatives examining relationships between individuals' health and well-being and their use of the outdoors. The goal was to develop a "holistic understanding of the connections between socio-economic and environmental factors that shape outdoors use, decision-making at the individual level, and health outcomes" and to stimulate further policy-relevant research." Another network, P_Infectious, is explicitly set "to catalyse interdisciplinary collaboration" and "develop novel methods", with the goal of improving the understanding of "how individuals may change their behaviour during infectious disease outbreaks". The CeEnviron Programme carried out "research on the causes, consequences and policy implications of global environmental change" focusing on "policy issues, including global warming, global biological diversity and institutional adaptation to global environmental change".
3. Where the main research driver was to make scientific contributions, **engagement with stakeholders (and through it problem orientation) was presented as a way to develop and improve such contributions**. Here the engagement with societal problems is instrumental to the theoretical concerns of the study. The UserICT programme, for instance, set as its central aim to drive "core academic results through deep user engagement." To generate user engagement the research needs to be relevant to user problems.

### 3.2. Cognitive distance

The "distance" between bodies of knowledge is very variable. Intuitively, one can guess that economics and social studies are cognitively closer than sociology and chemistry. Science mapping tools based on the citations that journal articles make to other journal articles support this view (Klavans and Boyack 2009; Rafols, Porter et al. 2010), as illustrated in Figure 1. This global map shows the distances among different fields of knowledge.

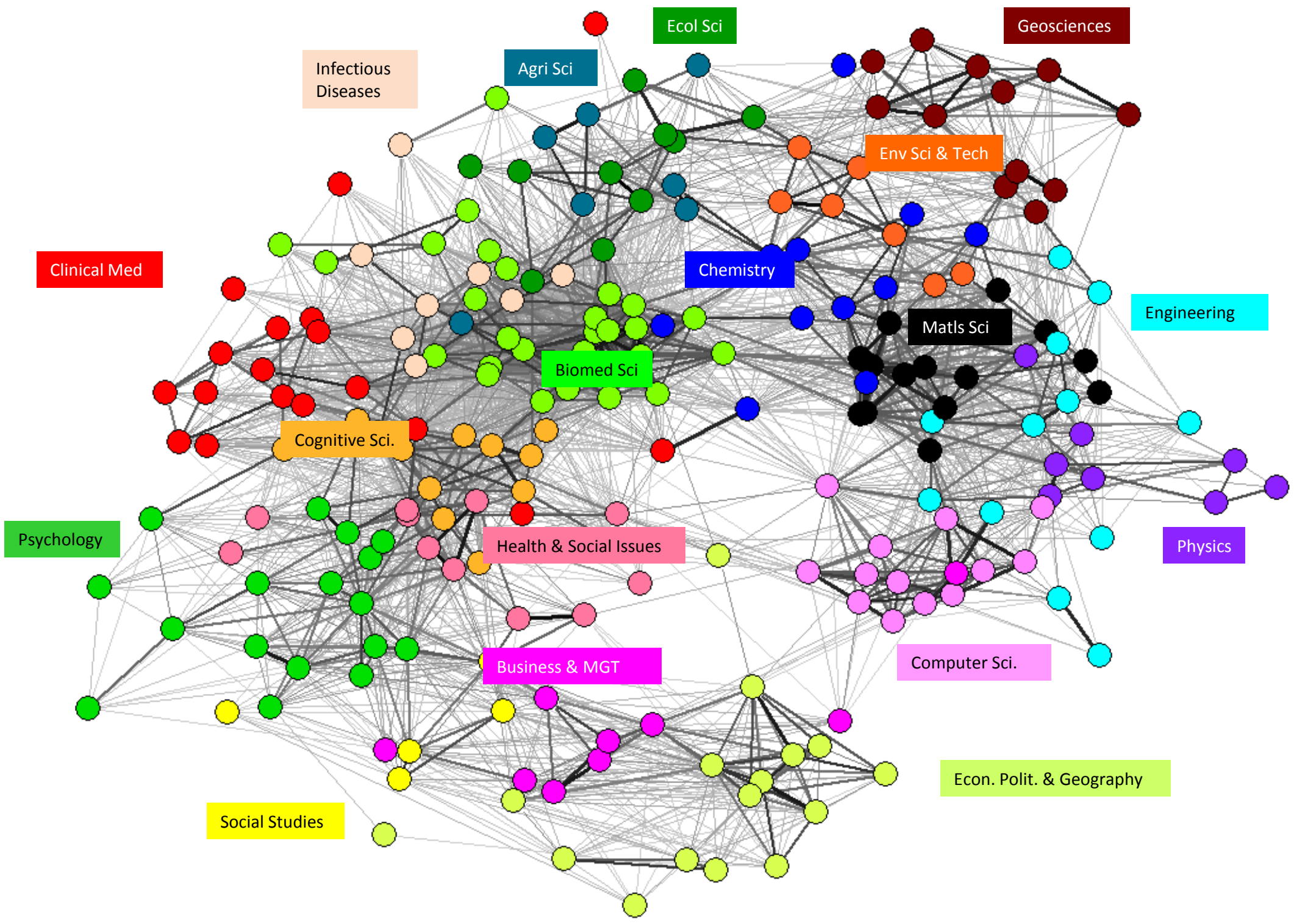

**Figure 1. The global map of science, illustrating the cognitive distance between disciplines. Source: Rafols, Porter and Leydesdorff (2010).**

The map can be used to "overlay" subsets of activity (a centre, a journal, an investment) and observe differences in the ways in which such subsets refer to different bodies of knowledge. For instance, Figures 2 and 3 show the location of publication and the citations between disciplinary fields for publication outputs generated by the NERC-funded QUEST programme and the ESRC-funded Innogen Centre respectively. What we can see is that while the QUEST programme involves neighbouring disciplines (in ecology and geology), Innogen involved distant disciplines such as biology, public health and management.

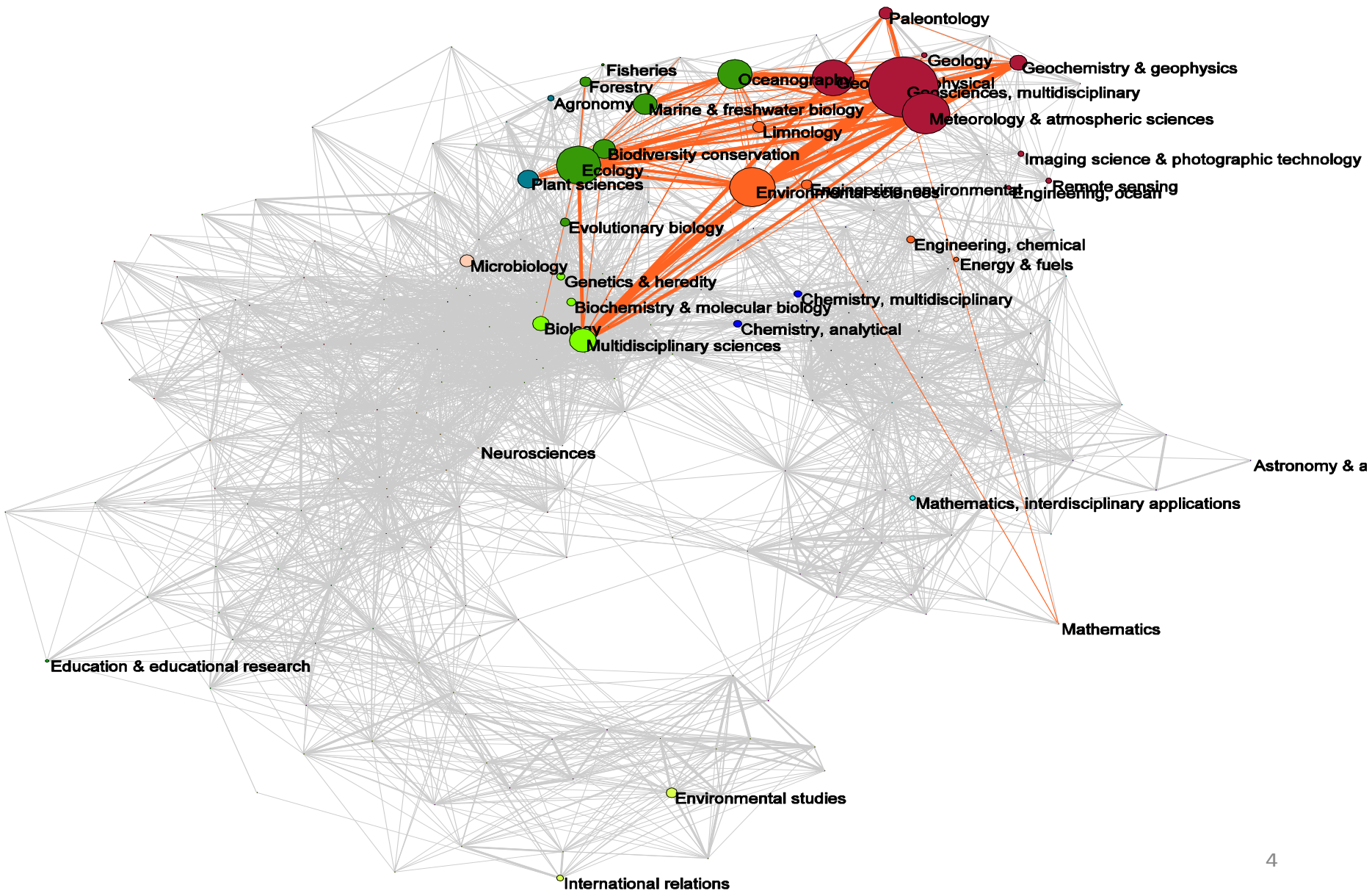

**Figure 2. Location of publication and citations between disciplinary fields (Web of Science Categories) for outputs of the QUEST programme (O'Hare and Rafols, 2011).**

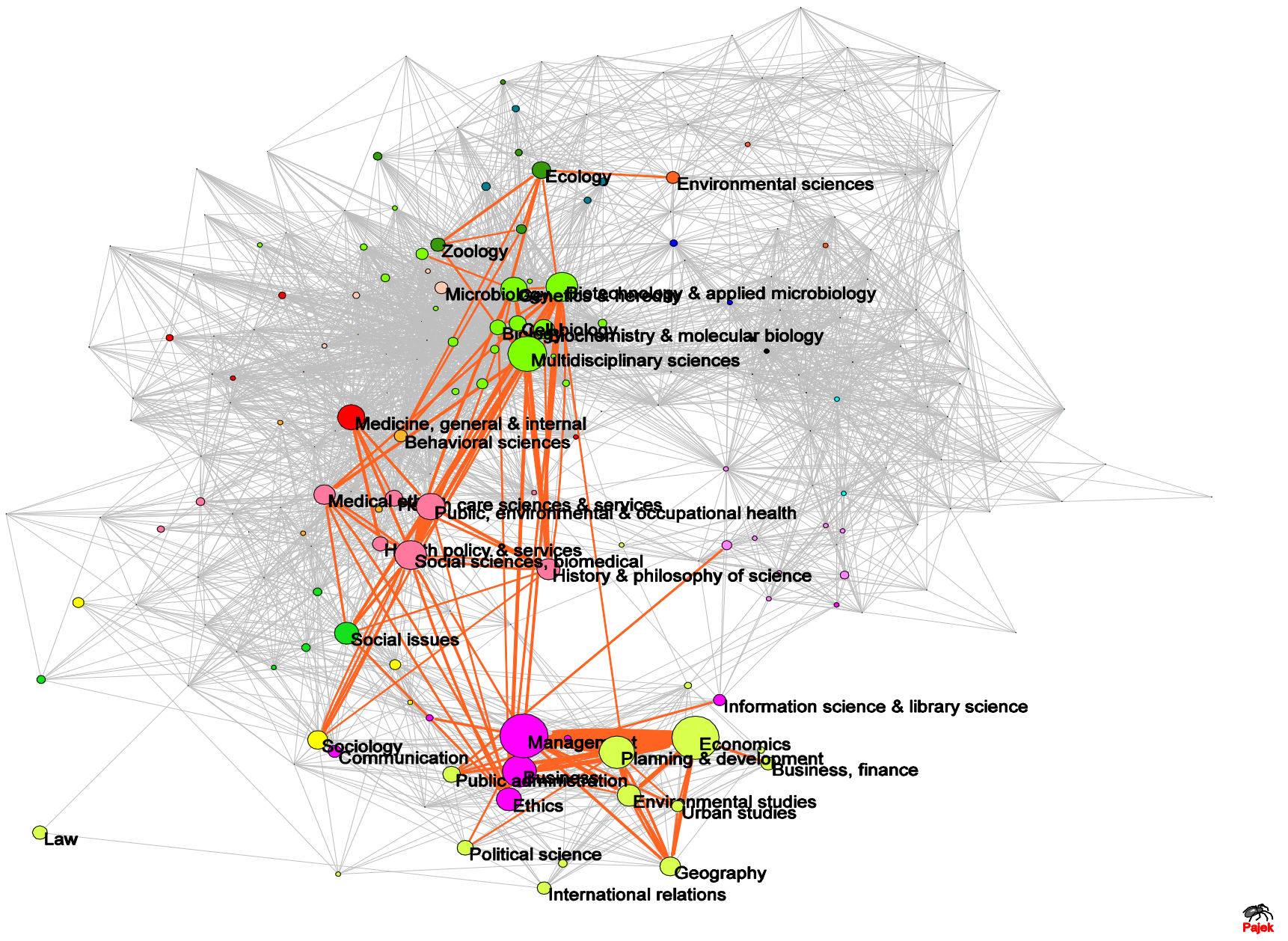

**Figure 3. Location of publication and citations between disciplinary fields (Web of Science Categories) for outputs of the Innogen Centre (Rafols and Costas, 2012)**

The generation of this type of maps was not within the remit of our study, but we could easily see in the cases we analysed that different investments involved different disciplines with varying cognitive distance. In some cases, projects combined close areas like P_HighEd in the social sciences, while in others they combined "distant" disciplines located in the social and life sciences and engineering. Combining distant bodies of knowledge is typical of investments addressing a specific and narrowly bounded problem. P_Water, for instance, used and contributed to a variety of disciplines as it published outputs in journals in areas like science studies, hydrological science, geography, environmental policy, social theory, and history and philosophy of science.

### 3.3. Extent and types of knowledge integration

The combination of different bodies of knowledge can result in different levels of knowledge integration. At one extreme multiple perspectives can be applied in parallel, leading to a richer understanding of a problem but without any integration of results or cross-feeding of insights. At the other end, a new integrated body of knowledge can emerge drawing on different disciplines but constituting a new form of knowledge that cannot be easily classed within any of them. In the review of ESRC projects, we have indeed found a wide range of outcomes regarding the extent of knowledge integration. P_Hydro claimed to have delivered "fundamental integration between natural science and socio-economic analyses" through the development of a modelling tool. P_Water also took a knowledge integration approach involving, in their own words, "'radical' collaborative ambitions both between natural/social scientists, and between these scientists and those affected by the environmental problem at issue – in our case flooding".

In other cases the collaboration did not revolve around integration, but there was a flow of knowledge through different disciplines. P_Sleep work packages were defined along disciplinary lines; there was one work package for psychologists, another for clinicians and nursing researchers, and another for occupational therapists. The findings of the work package conducted by clinicians and nursing researchers informed, to a great extent, the direction of the research conducted by the occupational therapists. The P_HighEd research team included Indian economists, and Brazilian and British researchers who were mainly sociologists. Here the researchers claimed to have brought "different expertise and approaches…challenging each other to ask different questions or use unfamiliar methods". This type of IDR research seeks to obtain "mutual stimulation" (Laudel 2001) from juxtaposing different perspectives.

A variation of this "multi-perspective" theme emerges when a perspective from a discipline is applied to a problem that has traditionally been addressed from different perspectives. IDR is seen in this case as a way of bringing new perspectives into research problems, without aiming at integration. P_Health, for instance, is presented as "an inter-disciplinary research project bringing together social sciences and philosophy", using "philosophical reasoning to begin addressing some foundational questions at the intersection of social epidemiology, development theory and practice, and social justice philosophy." In other words, the intention is to apply insights and approaches from one discipline (philosophy) to topics that may have been addressed from other disciplinary perspectives. The objective is to offer novel insights from disciplines that have not traditionally dealt with an issue, rather than integrate methods or theoretical insights developed from different disciplines.

### 3.4. Collaborative practices

A fourth distinction in the many different ways to conduct IDR is about the way in which IDR is practiced. IDR can be conducted by lone researchers or by groups (Palmer, 1999; Rafols, 2007). When involving groups of researchers, IDR may involve the formation of different teams according to the researchers' disciplines, each in charge of a clearly defined work package (P_Sleep), or there may be instances in which the researchers work jointly together (sharing work packages, working together in empirical research, or sharing in the interpretation and writing up of results). P_Water, for instance, deliberately sought to develop integrated perspectives and it did so by deploying collaborative teams working together.

## 4. The impact processes

### 4.1. Stakeholder engagement

It is well known that research impact is not only based on the transfer of research results to potential stakeholders and users, but can also occur through channels like training in new skills and capabilities or the progressive adoption of new concepts, framework or ideas. Further, impact may not only occur through a linear transfer of results to applications, but also through processes of collaboration between researchers and stakeholders that can be traced to the early research definition phases. Impact assessment literature is increasingly paying attention to these interactions (Molas-Gallart and Tang 2011; Spaapen and van Drooge 2011). They are not only a mechanism to convey better the results of research but a way to enhance the relevance as well as the definition and conduct of research through the involvement of diverse expertise in the research process.

Stakeholders can contribute to one or several of the following: (i) the definition of the research topic and/or problem; (ii) the definition of the research strategy and techniques; (iii) the conduct of the empirical research; and (iv) the analysis of the results. Since in making these contributions, stakeholders are bringing into the research process user knowledge, stakeholder engagement has been viewed by some researchers as a particular form of interdisciplinarity (Wagner, Roessner et al. 2009, p.16). Yet, the link between stakeholder involvement and IDR appears to be more present in countries with a tradition of consensus building or participatory democracy on certain issues, such as water resources management in the Netherlands (Merxk and van den Besselaar 2008) or sustainability in Switzerland (Scholz, Lang et al. 2006).

*Collaborative*

The relationship between stakeholders and researchers was evident in some of the investments analysed for this report. SustainEcon was deliberately set up as an IDR programme seeking stakeholder engagement from all its projects. SustainEcon's "Impact Report" concluded that the Programme "created a distinctive culture oriented toward addressing stakeholder issues" and that this was a factor contributing towards the impact of the programme and its projects. To achieve stakeholder involvement, SustainEcon deployed a variety of practices at the programme level: knowledge intermediaries, work-shadowing, visiting fellowships, final conference's Impact Awards,… One of the SustainEcon projects, P_Water, developed an "experimental collaborative research methodology - Competency Groups (CGs)" to bring together researchers from different disciplines and stakeholders directly affected by flooding events to study the problem the project was addressing. P_Water (one of the two projects we selected for interviews with researchers and stakeholders because of their reported stakeholder participation plan) exemplifies the role of stakeholder engagement from the start of the project. City councillors and the Flood Forum (made up of local residents) participated in it. The councillors collaborated with the research team for two years. These practitioners played a key role providing "flood" information to the research team and liaising with other local residents who had been initially reluctant to provide information to some of the researchers.

For the P_Sleep project, according to an interviewed researcher, a whole work package was dedicated to working directly with potential users of the research. This work package built on the findings of other work packages, which involved a range of disciplines. In it, care home residents and staff provided direct input into the prototypes of four products that the project had developed for use in care homes. Five selected care home residents tested the four prototypes and fed their input, via the staff, back to the research team. The night shift staff also monitored the use of these products and reported their observations to the research team. The feedback helped to inform the redesign of the prototypes. One of these products (a night tray organiser) was further refined for further development based on the input and is expected to be commercially marketed by the end of 2014. The trial period for this successful product involved five care home residents for a total of 26 days and was redesigned with each input. The trial period for all four concepts ran for about six months.

In the case of the UserICT programme, the promotion of this stakeholder engagement was laid out as a programme objective. The lead researcher from a participating project stated that the Programme had led to an attitudinal shift and that he would now begin to include stakeholders and seek their opinions from the very beginning of a project. Success in the promotion of stakeholder engagement was seen by the UserICT Executive Director as one of the achievements of the programme. In his final report he pointed out that UserICT projects had “41 collaborating organisations from private, public and voluntary sectors” and that all but one of these organisations and all but one of the 114 academic award holders “were new to this form of collaborative research.”

*Feedback*

A different form of interaction is when feedback from potential users is fed into the research. Here there are no permanent collaborative structures between researchers and stakeholders, but instead, feedback mechanisms are instituted to adapt aspects of the research to stakeholder inputs. P_Older included among its objectives to “iterate models according to insights and requests from policy makers where appropriate,” in addition to a more “linear” goal to “communicate results effectively to policy makers”. The project reported that several presentations, submissions and discussions were made to Parliament and in many workshops and events in other policy and technical fora. P_Older’s impact report referred to citations of their work in Parliamentary Committees and Government reports, as well as the provision of advice to Commissions.

*Transfer*

Other projects have deployed more limited engagement strategies with stakeholders, which mainly involved the transfer of research results. P_HighEd researchers used the relationships of its research partners in India and Brazil with national statistical offices and government departments to “have a regular dialogue” and make “presentations to policymakers”. Traditional dissemination tools like meetings, presentations and publications oriented to practitioners or potential beneficiaries were cited as the main ways in which relevance and impact was achieved by other projects (P_Infectious, P_Wellbeing, P_Crops).

P_Sleep, in addition to active engagement with users and practitioners in the course of their research as described above, also reported traditional dissemination work, with its researchers being “active in media dissemination (11 radio and 5 TV interviews and 35 newspaper/online items), targeting the general public and the scientific and clinical communities," plus the production of booklets, articles in practitioner magazines, training courses for people with sleeping problems and self-management for insomniacs among older people, and presentations.

The variety of ways in which stakeholders can be engaged in research projects has also been identified within a single programme. The evaluation of UserICT found that industrial partners played various roles within its projects, "including project leader (2 projects), potential user/tester of project technologies (10), contributor of project technologies (9) and advisor on commercial potential (5)." Such level of engagement was generated by the nature of the Programme and specifically its funding arrangements, rather than the type of research (IDR or otherwise) undertaken. UserICT evolved into a programme explicitly encouraging collaborations between academics and non-academics and involved the then-Department of Trade and Industry who co-funded eleven UserICT projects as collaborations with industry.

The projects reviewed suggest that involvement with industrial partners generated problems different from those encountered by projects with a policy orientation. The UserICT evaluation found that 16 of its projects involved industrial partners; yet, the unpredictability of the ICT sector, called for substantial efforts on behalf of the investigators and management team and concluded that "New structural approaches beyond existing 'knowledge transfer' measures are required to overcome this problem". Unpredictability as a problem emerging in dealings with industrial partners was also identified by P_Sleep. This project, as already noted above, developed products to support sleep but found the move from prototype to commercial products to be very difficult.

### 4.2.Types of impact

There are many different ways in which research activities can generate impacts. The investments analysed for this report are no exception. A UserICT evaluation we reviewed concluded that *within each project* in the programme one could find "three or more types of impacts", thus reflecting "the multi-layered complexity of the generation of impacts from research." It can be argued however that different ways of interacting with stakeholders are likely to generate different forms of impact. When the main form of interaction is through the dissemination of project results and such dissemination has reached its audiences it can be claimed that impact has occurred through informing the work of stakeholders (such as that found in P_Health). Another project claimed that the results had been used by government agencies "to advise policy" or by organisations as evidence in policy campaigns (P_Crops). Further evidence of impact can also be found in indicators showing that such results have been taken on board by the receivers of the information. Such indicators can be found, for instance, in the form of citations in official reports and documents (P_Wellbeing).

The notion of "impact" is also understood differently across the projects and evaluations we reviewed. In some cases the efforts to disseminate results, such as through publication of books, websites, meetings with stakeholders, and conference presentations were presented as evidence of impact (P_Health, Project P_Money). Yet, different rapporteurs assessed the value of dissemination outputs and activities *as forms of impact* very differently. For instance, one rapporteur was critical of the evidence of social-economic impact presented stating that "an MSc course [building around project results] is not exactly what research councils have in mind when considering 'economic and social impact'" while another for a different project estimated that a set of discussion papers and a website showed “good quality of impact”, because the papers were highly relevant and the website provided an “excellent” “information tool”.

Additional interactions after the project ended were taken by PIs as evidence of impact. P_Nature continued with the links with some of the stakeholders involved in the project through further work being contracted by the Forestry Commission and through membership of Government-sponsored evaluation groups and committees. Other project reports went a further step back, and considered the ability to engage in these interactions rather than the evidence of their continuation after the project, as an instance of impact. A report on UserICT considers “the ability to work across the academic/non-academic boundary” as a “capacity-building impact”.

Another result that has been reported as an impact is the effect of an investment on the future funding of a research line and the nature of such funding. SustainEcon has been acknowledged by senior officials at the Research Councils (RCs) to have had an impact on funders and has affected the way in which large investments funded by different RCs have been organised. The example put forward is that of ESPA (Ecosystems Services for Policy Alleviation) launched in 2009 with £40.5M of funding provided by NERC, ESRC, and DfID. An official at NERC interviewed for SustainEcon’s full “Impact Report” stated that “It’s had a big impact on us as funders and how we do things…SustainEcon has definitely influenced our thinking about how to approach interdisciplinarity and interface with social sciences—so it has been an invaluable learning experience—and on top of that, it has produced some excellent science”. The fact that several research programmes set up by NERC since then have had a socioeconomic dimension is attributed to SustainEcon’s influence.

Our interviews with stakeholders revealed direct traceable impacts on policy and practice. Three stakeholders who were interviewed for their views on the benefits of the P_Water project volunteered their views on the project’s impact. The two local councillors who participated in the project said:

> *“We thought that every area was surveyed but [the] Environment Agency just ‘mapped’ it and there were very few details. It was fortuitous that [the botanist] was involved in the project and she did surveys and then used the Overflow Model to see where debris dams (an anti-flooding measure recommended by the P_Water project) could be built to reduce flooding.” [*Councillor 1.]
>
> *“Defra*[1] *now seems to have learnt about building debris dams upstream. So it is important to know about cheaper options….the Environment Agency is now funding a project here….nothing would have happened [in the town] without the P_Water project.”* [Councillor 1.]
>
> *“The project helped enormously because [our town] does not have enough resources to manage floods…..and the residents kept telling Environment Agency about the P_Water project, which offered cheaper options, such as building debris dams and planting trees…the Agency became interested and funded a flood project [for the town].”* [Councillor 2]
>
> *“The Overflow Model is on a laptop, so it is easy to use and [it is] portable and so it was easy to demonstrate to us what to do and where to delay flooding….the legacy of this project gave us hope that floods could be managed….there is no magic cure for preventing flooding but at*

[1] Defra is the Department for the Environment, Food and Rural Affairs.

*least we now have a better understanding of how the river works and the Model showed us the spread of water across the flood plain."* [Councillor 2]

The above quotes illustrate the direct benefit and role of stakeholder engagement in the potential generation of impact. Yet an interview with a Government official revealed that P_Water also yielded indirect impacts. This official, a "stakeholder contact" named in the P_Water evaluation report, explained that he did not participate in the project, but that he had learnt about it through multiple sources, such as from colleagues who were aware of the project and from local authorities in Pickering with whom he deals with in the course of his work. This Government official went on to prepare a project proposal for Defra (and subsequently funded) in which one of the groups who participated in the P_Water project was invited to be a partner

"*because we were aware of the earlier work that the SustainEcon [P_Water] project had done with the local community in the catchment. [Our] University fed in their knowledge and the conclusions that arose out of SustainEcon [into the bid], but their main contribution to the Defra funded work was the application of their Overflow Model to help identify where to site Catchment Riparian Intervention Measures (CRIMs) for best flood risk reduction benefit."*

This Defra project was also made up of a range of participants with different knowledge, experience and skills, which included the local community "action groups," local authorities (town and district councils), academics (Durham University), the Environment Agency, the Forestry Commission, and national and regional organisations that were associated in various ways to land conservation and flooding issues. The collaboration approached developed by P_Water saw continuity in other projects to develop local solutions to flooding.

The review above and the interviews indicate that in the projects we have studied IDR involving close collaboration among researchers from several disciplines and stakeholders has emerged in projects seeking to develop *practical solutions* for specific problems that can be narrowly defined. These problems provide the focus both stakeholder engagement and IDR. Stakeholders bring to the project important knowledge of the local conditions in which the problem emerges. IDR is necessary to tackle the different types of practical difficulties that the problem poses. The link between the project and the impact is direct, and problem orientation, interdisciplinary work and stakeholder engagement are closely intertwined. Projects like P_Water and P_Sleep provide examples of how a **direct** connection between impact and research. Yet this "pathway to impact" is still compatible with more indirect forms of impact that can co-exist within the same project, as in the example provided by the Government official use of information from the P_Water project.

## 5. IDR modalities: a synthetic view

In the analysis carried out above we observe considerable variety even within a small group of initiatives funded by the same source, and these differences can often be nuanced. Table 1 below attempts to summarise the characteristics of all the investments analysed in this study. The classifications used in the table are very simple, and in trying to encapsulate the most significant features we lose the detail that a narrative description can offer. For instance, a project may be driven by both the wish to answer academic questions and propose solutions to applied problems; in the table below we make a sharp distinction between these different goals guided by the objective that emerges as central in the descriptions of the projects presented by the researchers themselves.

Similarly, there are many forms of socio-economic impact (Molas-Gallart, Tang et al. 2000), which we simplify here into two main types. Therefore, the Table does not aim to offer a complete view of the project analysed but instead to present a synthetic view of what we consider are their dominant characteristics.

In the column "**Drivers**" we present the main explicit factor that drives the research. We distinguish three main possibilities:

1. By "**specific problems**" we refer to projects aiming to provide specific solutions to problems that are clearly bounded and can be tackled directly by groups of stakeholders small enough to be able to participate directly in a research investment.
2. By "**challenge driven**" we mean projects that aim to contribute to the resolution of broad problems, with many different aspects, affecting society.
3. "**Scientific-Method**" when the main aim is to provide answers to questions of an academic nature. The main objective is to generate new knowledge through either the generation of new analytical frameworks, theories or research methods, or a combination of the above.

In the "**Cognitive distance**" column we differentiate two categories: "**SSH/short**" when the different disciplines that an investment combines or integrates belong to the social sciences and humanities, and "**SSH-STEM/long**" when a bridge is built to disciplines from the "hard" sciences, engineering or mathematics. There is more "cognitive distance" in the latter.

"**Knowledge integration**" refers to the way in which the different disciplines are combined within the investment. In the Table, we distinguish two main types: "**integrative**" when the aim is to provide new theories, methods and approaches that bring together knowledge that would have traditionally been classified in different disciplines. In these situations the theoretical outcome is difficult, when not impossible, to describe by resorting to a traditional disciplinary distinction and requires the definition of new descriptors. In contrast "**multi perspective**" refers to projects that bring together insights from different disciplines but that retain the disciplinary characteristics of each component of the research and its outcomes.

"**IDR Practice**" refers to the ways in which the researchers in an IDR project work. We distinguish two main possibilities:

1. "**Joint**" when researchers from different disciplines collaborate by working together in the same work packages and assignments
2. "**Specialised teams**" when researchers from different disciplines work separately in different assignments and work packages.

The "**Impact processes**" refer to the way researchers interact with stakeholders and the type of socio-economic impact that the investment has generated. These are drawn from the project reports and evaluations. In the table, we distinguish three main types of stakeholder interactions:

1. "**Collaborative**" when non-academic stakeholders are involved in the research process, helping or otherwise contributing to the study.
2. "**Feedback**" when information from stakeholders is incorporated into the research process without direct stakeholder collaboration with the researchers.

3. "**Transfer**" when the interactions with potential users and beneficiaries of the research involve *only* the communication of research results, either through formal or informal direct contact, or at arms' length through the broadcasting of research results.

Finally we only list two types of impact, although we are very aware of the very different ways in which research and its results can generate or help generate social and economic benefits. We also acknowledge that a simple classification does insufficient justice to the complexity of processes that help to generate the kinds of impacts. However, in the projects studied we were able to identify two main groups of impact: in some cases the documents analysed and our interviews provided evidence of the direct use by stakeholders or research results. Projects in this category ("**Use**") had yielded solutions with direct application. Other projects ("**Information**") had yielded data and information that stakeholders had received. The effects here are contributory, in the sense that there is evidence that they have been considered in and contributed to decision-making.

In some cases the documentation we reviewed did not offer enough information to enable an assessment of the categories to which the investment belonged ("**Not available"**). In other cases the documents interpreted the notions differently and therefore did not provide useful information for our analysis. This occurred mainly when researchers and reviewers considered the mere interaction with stakeholders or the dissemination of knowledge as evidence of impact; we have classified these situations as "**Not applicable**".

Even when using these very stylised categories,

Table 1 shows a variety of characteristics among the different projects. We cannot say that a specific form of impact or stakeholder interaction is necessarily associated with specific characteristics of IDR.

We can however identify two distinct, contrasting modalities consistent with different types of impact and impact pathways, which we would call long and short range interdisciplinarity. Loosely following the conceptual framework proposed by Boschma (2005) we propose that IDR can be analysed by thinking in terms of "distances" (short or long) across the analytical dimensions listed in Table 1.[2] Boschma proposed to look at five different dimensions: cognitive, social, organisational, institutional and geographical. In this analysis, we do not have a neat correspondence to all of them, but some of the columns are clearly related to some specific dimensions. The cognitive dimension is captured by cognitive distance ("SSH" has shorter distance of between areas of knowledge base) and knowledge integration ("integrative" shows shorter cognitive distance of interactions). IDR practice tells us about the organisational dimension --joint IDR has a "shorter" organisational distance than "specialised teams" . The institutional dimension is captured by stakeholder interactions, since "collaborative" arrangements mean that institutional contexts that are distant (e.g. university and industry) are brought into closer contact. Hence, we distinguish two modalities:

1. A **long range IDR modality** in which a set of cognitively distant disciplines are integrated through the joint work of interdisciplinary teams to address clearly defined specific problems. To do this successfully the initiative draws on the active collaboration of stakeholders and generates solutions that can be directly applied to the problematic situation, often as part of the research initiative itself. In the investments we analysed the SustainEcon programme and several of its projects provide examples of this modality.

[2] We have developed a related approach for the analysis of translational research at Molas-Gallart et al. (2014).

2. A **short range IDR modality** in which the contributions from different, cognitively close perspectives are harnessed to address a broadly defined social or economic problem; in other words, a societal challenge. The researchers work within their areas of expertise and provide a diversity of perspective. Coherence is maintained by dealing only with cognitively close disciplines (in our case in the social sciences and humanities only). The research results provide novel multidisciplinary perspectives that can then be conveyed to the broad stakeholder communities interested in the societal challenge being addressed. The way such communities use this information can be difficult to trace, but they are likely to use this information in combination with information provided from other sources. Several of the projects analysed for this study conform to this modality (P_HighEd, P_Older, P_Health)

These modalities are contrasting in that they differ in each and every one of their characteristics (the columns in

Table 1) and they can be helpful in defining the traits of IDR that a specific investment wants to address. In reality, however, projects will often combine different characteristics of these two contrasting modalities, as it was the case in many of the investments studied for this report (CeEnviron, P_Crops, P_Wellbeing,…). In general the long range IDR puts into contact spaces that are distant and makes an effort to link them, whilst the short range IDR operates in short cognitive distances and does not put so much attention to fostering links across actors that are distant in organisational or institutional terms.

**Table 1 IDR Modalities and Impact: a synthetic view of the projects analysed**

| Investment | Type | Drivers | Cognitive distance | Knowledge Integration | IDR Practice | Stakeholder interactions | Types of impact |
|---|---|---|---|---|---|---|---|
| SustainEcon | Programme | Problem-driven | SSH-STEM / long | Integrative | Joint | Collaborative | Use |
| UserICT | Programme | Scientific-Method | SSH-STEM / long | Multi-perspective | Joint | Collaborative | Use |
| SustainTech | Programme | Challenge-driven | SSH / short | Multi-perspective | Specialised teams | Transfer | Information |
| CeEnviron | Centre | Challenge-driven | SSH-STEM / long | Not available | Joint | Transfer | Information |
| P_Health | Project | Challenge-driven | SSH / short | Multi-perspective | Not available | Transfer | Not applicable |
| P_Money | Project | Problem-driven | SSH / short | Not available | Not available | Transfer | Not applicable |
| P_Water | Project (part of the SustainEcon). | Problem-driven | SSH-STEM / long | Integrative | Joint | Collaborative | Use |
| P_Crops | Project (part of the SustainEcon) | Problem-driven | SSH-STEM / long | Multi-perspective | Not available | Transfer | Information |
| P_Hydro | Project (part of the SustainEcon) | Problem-driven | SSH-STEM / long | Integrative | Not available | Collaborative | Use |
| P_HighEd | Project | Challenge-driven | SSH / short | Multi-perspective | Specialised teams | Transfer | Information |
| P_Older | Project | Challenge-driven | SSH / short | Multi-perspective | Not available | Transfer | Information |
| P_Sleep | Project | Problem-driven | SSH-STEM / long | Multi-perspective | Specialised teams | Collaborative | Use |
| P_Infectious | Project | Challenge-driven | SSH-STEM / long | Multi-perspective | Specialised teams | Transfer | Information |
| P_Wellbeing | Project. | Scientific-Method | SSH-STEM / long | Multi-perspective | Specialised teams | Transfer | Information |
| P_nature | Project | Challenge-driven | SSH-STEM / long | Multi-perspective | Specialised teams | Not available | Not applicable |

## 6. Perceptions on IDR and Impact

### 6.1. How stakeholders see the effects of IDR

The way in which stakeholders assess the effects of IDR will obviously depend on the type of IDR and the interactions to which they have been exposed. A "Policy and Practice Impact Case Study" of CeEnviron concluded that "end users" saw the interdisciplinary nature of the programme as beneficial as it "allowed CeEnviron to look at issues more holistically and consider wider interactions and policy implications. This enabled policy makers to act on CeEnviron research with greater confidence. It also added to the 'one stop shop' effect, allowing CeEnviron to offer a more complete (and possibly unique) service to policy makers." Researchers involved in the Programme also considered that the interdisciplinary approach had had benefits for the outlook of the staff, and one member suggested that it had prevented "CeEnviron from becoming too academic".

Views on the benefit of IDR were also gathered in the interviews with three P_WATER stakeholders. Local councillor 1 said "that without interdisciplinary research, [one] can't look at problems from all angles. Interdisciplinary research is 'value for money.'" Local councillor 2 echoed a similar view: "interdisciplinary research is most important. I am a great believer in interdisciplinary research because you can look at problems from different angles and come up with solutions. Single disciplinary research often has "a silo mentality."

The local councillors also added that because the research team had addressed the flooding problem from different perspectives and had involved the collaboration and participation of local communities and residents, these factors helped to encourage input and feedback to the project team. These sentiments were corroborated by the botanist researcher (again, a local resident, and therefore both a researcher in the project and a stakeholder) who claimed that by including different disciplines and local residents into the project, the understanding of the problem (flooding) was evident and the trust issue on the part of local residents was resolved, thereby leading to productive interactions with the local community.

The Government official explained that societal problems

> *"need different areas of expertise for problem-oriented solutions. This is crucial. I learnt this the hard way….when I did not include a key stakeholder organisation at the outset [of the project]. I underestimated the constraints to problem-solving when I did not include this key stakeholder which had special 'scientific' interests and knowledge in the area in which the project research was going to be conducted. This stakeholder organisation, upon hearing about the project, posed several problems and resistance to the project. If this situation was left unaddressed, the project would have run into difficulties. I quickly invited this organisation to be part of the project team."*

This official also added that the involvement of key stakeholders, especially and including local communities, is instrumental to good input and feedback for the project. Dissemination activities, according to him, are also crucial for helping to generate feedback and impact especially when the project is also dependant on the cooperation and participation of local residents. He noted that for his project he had a dedicated communications person who regularly emailed local residents about project events and developments, held "open evenings" to discuss the project and solicit views, knowledge and experience, and published bulletins and newsletters. These activities were not only to inform the local communities but also to gain their trust so that they would engage actively in the project. Time-consuming as this process was, this official claimed that the local communities eventually took over the dissemination of the project's progress and findings, when funding for the full-time communications person ended two years into the project.

### 6.2. How researchers see the effects of IDR

Both project documents and interviews show that researchers involved in the interdisciplinary projects we have analysed saw IDR as having clear and identifiable benefits. Perhaps most explicit in this respect was the SustainEcon full Impact Report. The report presents the results of an interview programme with researchers and stakeholders. All (100%) Programme Stakeholder, 91.7% Researcher and 76.4% Project Stakeholder respondents agreed that SustainEcon's emphasis on interdisciplinarity enhanced the capacity of SustainEcon researchers "to deliver usefully integrated understanding relevant to stakeholder problems". Nearly all (94.5%) Researcher respondents also agreed that "SustainEcon's emphasis on interdisciplinarity has enhanced the capacity of SustainEcon researchers to engage with stakeholders having different perspectives".

The researches we interviewed for this project were also keen to stress the importance of IDR. A researcher each in P_Water and P_Sleep attested to the need for IDR teams when addressing societal problems. They argued that problem-orientation research dealing with social problems requiring practical solutions benefit greatly from the knowledge, experience and skills of different academic perspectives and training. Both also underscored the role of participation of stakeholders because not only their knowledge, experience and awareness of the problems provide invaluable input to the research but also because they encourage the participation of the "local folks" (as in the case of P_Water) who often are reluctant to be forthcoming with their input, either because of the "lack of trust" or "simply shyness."

The P_Sleep researcher, as already noted above, confirmed that her research, which focused on user aspects, was informed by the findings of the other work packages and therefore by the contributions from a variety of disciplines. In particular these helped her structure her research and identify the issues she should focus on.

Both researchers also affirmed that at least in the projects they were involved in, IDR helped to foster user feedback because it enabled the researchers to have a good understanding of the various aspects of the problem they were investigating. This in turn provided the research teams with a firmer footing to present their research aims in a "sensitive" and knowledgeable fashion to users, which won over their confidence and their willingness "to be the experiment" (to quote one local councillor). Both researchers also believed that by including a range of perspectives and skills, interdisciplinary teams can tackle better the challenges posed by problems calling for practical solutions.

### 6.3 Summary of views on IDR

To summarise, the views on IDR from project researchers and stakeholders were overwhelmingly positive. There were two main reasons for this positive outlook. First there is a broad consensus drawn from both the interviewees and project documentation that IDR, by offering different perspectives, helps address complex problems. Understanding the different angles of a problem is a precondition for its solution and by providing a more holistic view of a situation IDR can provide, to quote one report, "a better service". Perhaps less expected was the relationship that interviewees drew between IDR and stakeholder engagement; IDR facilitated engagement because it helped researchers to provide a more wide-ranging view of the problems for which they were seeking stakeholder support, and in this way to win their trust.

The above testimonies by the stakeholders illustrate how user feedback can be fostered if stakeholders can trust the researchers, be convinced of the aims of the research, understand that the project is being approached from "different angles," know that their input and feedback will be listened to and importantly, that the stakeholders are involved in the project itself. This combination of factors facilitates greater opportunities for generating impacts. Under these conditions, the relationship between IDR and stakeholder engagement can be mutually reinforcing. An IDR approach can benefit from stakeholder insights as part of the research process, and the views of stakeholders become one of the perspectives that IDRs consider. In addition, IDR can also facilitate user engagement by helping researchers gain the trust of stakeholders. It must be noted, however, that we have only found evidence of this mutually reinforcing relationship in the long range modality IDR.

## 7. Conclusions

Academic disciplines establish boundaries and set up epistemological rules for scientific work addressing specific subjects. Disciplinary research is generally triggered by scientific problems conceived within dominant paradigms and disciplinary perspectives. Thus, disciplinary research accumulates knowledge that matters for the research agendas of the specific research communities associated with a discipline. The knowledge generated within disciplinary boundaries may not fit the nature of the practical problems that individuals and societies encounter. These problems do not emerge from the scientific work carried out within a discipline and, therefore, addressing them is likely to require knowledge combinations that transcend disciplinary boundaries.

The research activities we reviewed for this study were mostly driven by a concern to address societal problems; such concern could develop in parallel with the goal of making valuable theoretical contributions, but theory alone did not drive the research. In the projects reviewed we observed that the relationship between relevance and interdisciplinarity is complex: all projects in principle addressed social and economic problems and issues but understood interdisciplinarity and impact in different ways. However, in the investments we have reviewed we identified two contrasting IDR modalities linked with different impacts and impact pathways. In the modality that we have defined as "long range", the demands of addressing specific problems were associated with approaches that integrated cognitively distant disciplines to develop solutions for these problems (such as developing strategies to deal with flooding in P_Water). In the "short range" modality different, but cognitively close disciplinary perspectives are combined without seeking integration to address societal challenges (for instance, the study of the social determinants of ill-health in P_Health). In this latter situation, impact was typically sought through the dissemination of research results.

It is important to note that if we understand impact as evidence that stakeholders are doing things they were not doing before the research, the instances of impact we found were related to the use of research and research results in the “long range” IDR modality. As we have described above, this modality is oriented to the solution of clearly bounded problems, for which there is a need to integrate contributions from different disciplines in active collaboration with stakeholders. In these situations, stakeholders were involved in the processes that led to the identification and definition of the problems to be addressed and in the research itself. This modality of IDR is well justified: problem orientation requires awareness and knowledge of the societal problems to be addressed and such awareness and knowledge needs to be developed through interactions with practitioners.

There is therefore a link between the modalities of IDR we have encountered and their potential for direct, traceable impact. This link, is underpinned, first, by a specific approach to problem orientation that occurs when precise, clearly bounded social problems are explicitly identified as needing solutions. Second, as we have discussed, some of the initiatives studied for this report explicitly reported evidence of a link between IDR and impact taking place through a strong stakeholder engagement. In short, IDR seeking to develop solutions by involving stakeholders in close iterative interaction appears to be closely associated with our ability to trace examples of research impact attributable to the research investment.

Figure 4 summarises this argument. In the "long range" modality, IDR is related to impact because it supports and it is in turn shaped by problem-orientation and stakeholder engagement, and it is through problem orientation and stakeholder engagement that impact occurs. Obviously, this is not the only way in which IDR can be associated with social impact. In the “short range” modality, research is also driven by the desire to tackle social and economic issues, but these are defined broadly as challenges that society faces and that pose a variety of specific problems. The relationship with stakeholders is also very different in this modality: they are not directly involved in the definition and conduct of the research but as potential users and beneficiaries of its results.

We have to stress that this is an stylised analysis: many of the investments we analysed display characteristics reflecting a mix of elements of the “short range” and “long range” modalities. Therefore, the IDR label covers very different ways of conducting research that are conducive to different ways of generating impact. We cannot say that IDR is associated with a distinct type of impact mechanism. Terms such as “impact of IDR” can therefore be misleading as they may suggest that practices are more homogeneous than they are in reality.[3]

IDR is too diverse to be a useful category for science policy management. Even within the narrowly bounded scope of this analysis, which focused only on research funded by a specific funding body targeting the economic and social sciences within a single country, we found a wide variety of ways of understanding what IDR and impact are, of conducting research and of seeking social impact.

[3] An researcher affiliated in P_Water has evaluated P_Water team's experiences of interdisciplinary working and concluded that “the conventions of interdisciplinarity should be considered as ideal types, which do not reflect the actual mess of interdisciplinary research in practice.”

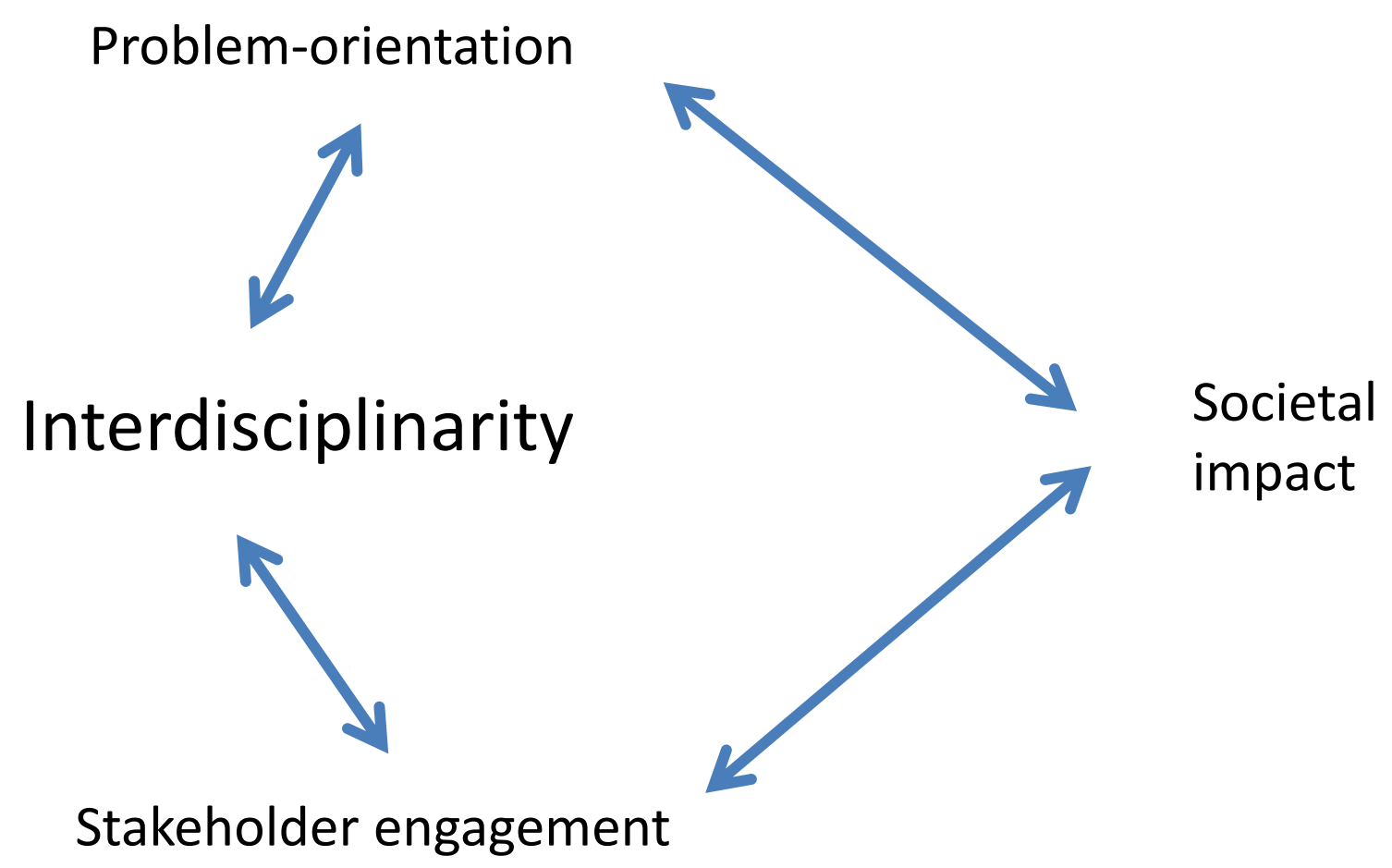

**Figure 4. Long Range IDR: The relationship between interdisciplinarity and societal impact.**

**8. Policy implications**

An important observation of this study is that there is no single interdisciplinary approach that will help to generate impact. In addition, IDR is neither a sufficient nor a necessary condition for societal relevance and impact: it can also emerge as a response to purely academic questions and we can think of societal problems that can be tackled within the confines of traditionally bounded disciplines. Therefore it is not meaningful to speak of IDR as a well defined set of research practices sharing a common way of generating societal benefits and to draw recommendations that are applicable, in general, to all IDR modalities. For instance, an IDR modality conducive to impact involves the integration of capacities and knowledge from different disciplines to tackle specific, clearly bounded problems in collaboration with stakeholders. Of course this is not the only form of IDR conducive to impact generation, neither is the combination of problem orientation and stakeholder engagement unique to IDR. Yet, most societal problems cannot be tackled using the knowledge generated by a single academic discipline. When this occurs IDR projects can engage stakeholder communities more easily as they offer a breadth of skills and capacities that can be attractive to stakeholders who are aware of the characteristics and complexities of the problems faced.

When developing policies on IDR, funding agencies should be clear about the modality of IDR it wishes to support and the reasons for doing so. Supporting IDR "in the abstract" will lead to a lack of focus and clarity and will therefore make evaluation of the intervention very difficult. Therefore the following key question needs to be addressed:

1. What modality of IDR does the funding agency wish to support? IDR can be interpreted in many different ways, both by researchers, proposal reviewers and project evaluators.
2. Does the intervention seek views from different disciplines in the social sciences and humanities to shed additional light on a broad societal challenge?

3. Does it aim to bringing together distant disciplines to solve specific problems? The latter may require cross-agency or council collaboration and the inclusion of non-academic stakeholders in research design and performance.

The term IDR should also be unpacked to describe in more detail the approach envisaged:

1. the type of problems to be addressed;
2. the variety and distance between disciplines to bridge;
3. the form of collaboration;
4. the extent and form of stakeholder involvement.

The criteria for ex-ante and ex-post evaluation and the composition of the reviewing panels should then be designed in accordance with these characteristics. Further, if stakeholder engagement is seen as an important aspect of the modality of IDR research to be supported, the application forms will have to be adapted accordingly. For instance, "pathways to impact" statements cannot be seen as add-ons dealing with the dissemination of research results, but as part and parcel of the scientific methodology used. If such changes are put in place, then they need to be accorded due attention during the proposals' review process.

Clarifying the forms of collaboration that an intervention seeks to support will also help in its ex-post evaluation. The different ways in which interdisciplinarity can be approached mean that it can be assessed differently using different criteria. We found signs of this problem in the documents we reviewed. One project rapporteur complained that the project showed "little understanding" of what "integration" and "interdisciplinarity" meant and that it had only achieved a set of individually "useful" outputs that reflected little interdisciplinary integration. Yet another rapporteur for the same project seemed satisfied that the project had examined a variety of issues ranging from the assessment of public attitudes to the analysis of biodiversity. Obviously, the two reviewers had different views on the meaning of interdisciplinarity and on the importance of interdisciplinarity to the achievement of project objectives.

We believe there is wide scope for these difficulties to reappear, unless investments and projects have a clear understanding of the modality of IDR being supported. Diversity means we cannot provide recommendations on how to maximise the benefits of IDR and address its challenges that are equally applicable to all possible IDR modalities. Different modalities will require different strategies. Long range approaches require problem-orientation to be clear, explicit and narrowly targeted and stakeholder engagement to be intensive throughout the research. Short range modalities can tackle broader societal issues aiming to contribute to a broad policy debate through their research results. These contributions will be richer if they come from different disciplines but the multiple perspectives may not need to be integrated into a single theoretical body. Ex-ante and ex-post evaluation need to be guided by clear criteria based on the modality of IDR that is being pursued.

**Acknowledgements**

This study exploratory study is a slightly modified and abridged version of a report submitted to the Economic and Social Research Council (ESRC) of the UK. Available at: http://www.esrc.ac.uk/_images/esrc-interdisciplinary-research-report_tcm8-32266.pdf (Dec. 2014)

**Annex 1: Anonymised list of investments analysed**

Research programmes

**SustainEcon:** Programme on sustainable economy, with a strong emphasis on agriculture and rural development. 10 years.

**SustainTech:** Programme on technologies for sustainability. 5 years.

**UserICT:** Programme on information and communication technologies with a focus on general public usage. 7 years.

Research centres

***CeEnviron:*** Centre on environmental studies at a global scale. 17 years.

Research projects

**P_Crops:** Investigation on the social consequences of certain types of crops. 3 years.

**P_Health**: Project that aimed to integrate quantitative health research and development studies approaches to social justice. 3 years.

**P_HighEd**: Project on the higher education in developing countries. 5 years.

**P_Hydro**: Study of hydrology management. 5 years.

**P_Infectious:** Project on behavioural attitudes to infectious diseases. 2 years.

**P_Money:** Project that investigated the management of money under betting stress. 3 years.

**P_Nature:** Study of the effects of enjoying the natural world in health. 2 years.

**P_Older:** Modelling study on needs of older population. 4 years.

**P_Sleep:** Project on sleep in older population. 5 years.

**P_Water:** Project that studied flood management. 4 years.

**P_Wellbeing:** Project on daily routines for well-being and the environment. 2 years.